\documentclass[sigconf]{acmart}
\usepackage[utf8]{inputenc}
\usepackage[labelfont=bf]{caption}
\usepackage[justification=centering]{caption}
\usepackage{esvect}
\usepackage{amsmath}

\newtheorem{definition}{Definition}

\date{April 2020}

\usepackage{graphicx}

\setcopyright{acmcopyright}
\copyrightyear{2020}
\acmYear{2020}
\acmDOI{10.1145/1122445.1122456}

\acmConference[AsiaCCS'20]{AsiaCCS'20: The 15th ACM ASIA Conference on Computer and Communications Security}{October 05--09, 2020}{Taipei, Taiwan}
\acmBooktitle{AsiaCCS'20: The 15th ACM ASIA Conference on Computer and Communications Security,
  October 05--09, 2020, Taipei, Taiwan}
\acmPrice{15.00}
\acmISBN{978-1-4503-XXXX-X/18/06}

\begin{document}

\title{Blockchain-Based Differential Privacy Cost Management System}

\author[1]{Leong Mei Han}
\affiliation{
\institution{Nanyang Technological University}
\country{Singapore}}
\email{MLEONG008@e.ntu.edu.sg}
\authornote{Both authors contributed equally to the paper. The order of names is alphabetical.}

\author[2]{Yang Zhao}
\affiliation{
\institution{Nanyang Technological University}
\country{Singapore}}
\email{S180049@e.ntu.edu.sg}
\authornote{Corresponding author} \authornotemark[1]

\author[3]{Jun Zhao}
\affiliation{
\institution{Nanyang Technological University}
\country{Singapore}}
\email{junzhao@ntu.edu.sg}

\begin{abstract}
Privacy preservation is a big concern for various sectors. To protect individual user data, one emerging technology is differential privacy. However, it still has limitations for datasets with frequent queries, such as the fast accumulation of privacy cost. To tackle this limitation, this paper explores the integration of a secured decentralised ledger, blockchain. Blockchain will be able to keep track of all noisy responses generated with differential privacy algorithm and allow for certain queries to reuse old responses. In this paper, a demo of a proposed blockchain-based privacy management system is designed as an interactive decentralised web application (DApp). The demo created illustrates that leveraging on blockchain will allow the total privacy cost accumulated to decrease significantly.
\end{abstract}



\maketitle
\thispagestyle{fancy}
\pagestyle{fancy}
\lhead{This paper appears in ACM ASIA Conference on Computer and Communications Security (ACM \textbf{ASIACCS}) 2020.\\ Please feel free to contact us for questions or remarks.}
\cfoot{\thepage}
\renewcommand{\headrulewidth}{0.4pt}
\renewcommand{\footrulewidth}{0pt}

\section{Introduction}
In today’s digital age, Internet has given us the potential to collect and access various kinds of information easily. Individuals are willing to give away personal information for online convenience \cite{craig2011privacy}. This vast amount of data about individuals is being constantly stored in various databases. When analysed efficiently, data can translate into meaningful information about the individual’s behaviour. This leads to personal data being the new currency of the digital age and is sought after by every industry and government. Despite the massive benefits that data analysis can bring, improper handling of sensitive data usually leads to more information being revealed than intended. One attempt to preserve user privacy was to release anonymised or aggregated data. However, it is proven that records can be de-anonymised when combined with other data sources. This happens when attackers can accurately match the anonymised database to another non-anonymised database \cite{narayanan2008robust}.

The introduction of differential privacy has brought us closer to achieving the goal of preserving personal privacy while still revealing meaningful information about datasets. The brief idea behind differential privacy is to incorporate some noise to the result of an output such that it does not change significantly with or without the addition of a single input in the dataset.

($\epsilon,\delta$)-Differential Privacy \cite{dwork2006calibrating} is defined as:
\begin{definition}
A randomized algorithm $Y$ satisfies ($\epsilon,\delta$)-Differential Privacy if, for any two neighboring datasets differing by one record ($D$ and $D^{'}$) and all subsets S of the output range, it generates a randomized output such that:
$Pr[Y(D)\in S] \leq e^{\epsilon}Pr[Y(D^{'})\in S]+\delta$, where the probability space is over the coin flips of the randomized algorithm $Y$.
\end{definition}
Traditionally, every query processed with differential privacy will generate a privacy cost. This privacy cost accumulates even if it is the same query. It brings about an issue where the privacy cost may exceed the privacy budget, leading to a greater percentage of privacy leakage \cite{cuppens2019optimal, jia2019database}.

The rise of blockchain technology gives rise to a possible solution to the above problem of privacy budget exhaustion. By using blockchain’s key advantages of decentralisation, tamper-proofing and traceability, it provides a distributed, trusted platform of peer-to-peer network to store information regarding the queries processed with differential privacy \cite{sultan2018conceptualizing}. These transaction data can then be retrieved later to be processed for possible reuse of previously generated noisy answer. This reusing of old noisy answer will be highlighted in the subsequent web DApp demonstration. With the proposed blockchain-based privacy management system, the total privacy cost incurred will be significantly reduced, catering for datasets with frequent queries such as medical record datasets.

\section{Related Works}

Zyskind~\emph{et al.} \cite{zyskind2015decentralizing} tackle the issue about privacy when using third-party mobile platform by suggesting the combination of blockchain with off-blockchain storage to create a personal data management platform for increased privacy. This allows users to have ownership and control over their data without requiring to trust any third-party. Kosba~\emph{et al.} \cite{kosba2016hawk} proposes a framework for building privacy-preserving smart contract. The proposed framework, Hawk, allows any programmer to easily write a program that implements a cryptographic protocol between blockchain and the user. This cryptographic protocol includes using authenticated data structures and zero-knowledge proofs for added security.

The above studies among others focused on identity privacy between the blockchain and the users, while trusting the anonymity of the blockchain. However, there is a lack of literature which focuses not only on the privacy of blockchain, but also the privacy protection for the database itself. Research on reusing noisy answers for privacy protection has also been done previously. 

Xiao~\emph{et al.} \cite{xiao2011ireduct} proposed a differentially private algorithm that correlates Laplace noise added to different query results for improved data utility. However, in this paper, Gaussian noise is used over Laplace noise for an easier privacy analysis of multiple query types. The sum of independent Laplace random variables does not follow a Laplace distribution. On the other hand, the sum of independent Gaussian random variables still follows the Gaussian distribution. As such, using Gaussian noise will allow the algorithm to handle queries of different types.

\section{Demo Overview}

\begin{figure}[h!]
\centering
\frame{\includegraphics[scale=1, height=90pt]{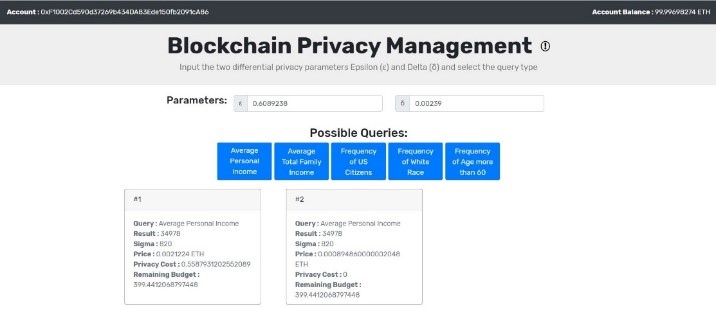}}
\caption{Screenshot of blockchain-based privacy management system demo.}
\captionsetup{justification=centering}
\label{fig:blockchaindemo}
\end{figure}

To illustrate the proposed blockchain-based privacy management system, a decentralised web application is created. This demo simulates how the system implements differential privacy algorithm while tracking and reducing the privacy cost incurred. Fig. \ref{fig:blockchaindemo} shows the overall user interface of the blockchain-based demo. 

\subsection{Ethereum account information}

\begin{figure}[h!]
\centering
\frame{\includegraphics[scale=1]{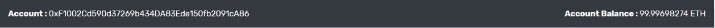}}
\caption{Displaying Ethereum account information of user.}
\captionsetup{justification=centering}
\label{fig:ethereumaccount}
\end{figure}

Ethereum account information of the user is displayed on a bar fixed at the top of the page once the user granted permission to the demo. The demo gets the account information (wallet address and balance inside it) from MetaMask extension of the browser. When connecting with MetaMask, a pop-up that is managed by MetaMask will appear asking users for their permission to access the account. Account information will only be displayed after permission is granted. 

\subsection{Defining differential privacy parameters}

\begin{figure}[h!]
\centering
\frame{\includegraphics[scale=1]{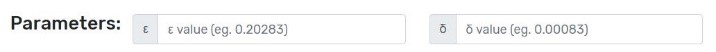}}
\caption{Capturing user’s desired differential privacy parameters.}
\captionsetup{justification=centering}
\label{fig:parameters}
\end{figure}

In the demo, users will be able to specify parameters used for differential privacy algorithm ($\epsilon{}$ and $\delta{}$) using the input bar shown in Fig. \ref{fig:parameters}. $\epsilon$ value is used to determine how strict the level of privacy is. The smaller the $\epsilon$ value, the better the privacy preservation. $\delta{}$ defines the level of relaxation of the $\epsilon{}$-differential privacy notion. 

\subsection{Queries selection}

\begin{figure}[h!]
\centering
\includegraphics[scale=1,width=\linewidth]{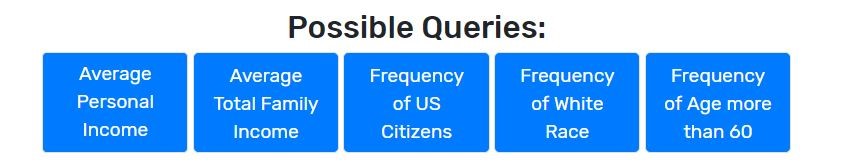}
\caption{Types of queries supported.}
\captionsetup{justification=centering}
\label{fig:queries}
\end{figure}

Users can select the query type that they wish to inquire about by pressing on the buttons created. In the backend, each button is linked to a sensitivity level that is pre-calculated and assigned to the button according to its query type. Differential privacy algorithm scales the noise generated with the sensitivity of the query function. This sensitivity level is the maximum distance between the true query results for any two neighboring datasets that differ by one record. The calculation of the sensitivity level follows: 
$$\Delta Q= max_{neighboring}  D, D^{'}\|Q(D)-Q(D^{'})\|_{2},$$
where $\Delta Q$ is the sensitivity level of the query Q, and D and $D^{'}$ are neighboring datasets that differs by one record.

\subsection{Reuse of previous Gaussian noise}

\begin{figure}[h!]
\centering
\frame{\includegraphics[scale=0.55]{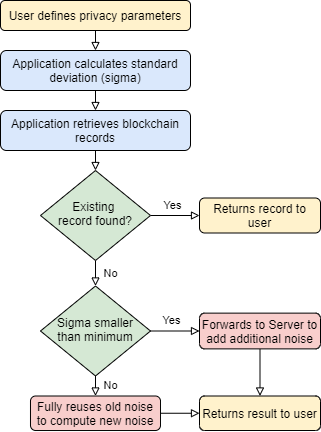}}
\caption{Workflow of noise reuse.}
\captionsetup{justification=centering}
\label{fig:noisereuse}
\end{figure}

The algorithm is designed to run on the comparison of standard deviation and follows the general workflow in Fig. \ref{fig:noisereuse}.

Based on the privacy parameters from the user, a $\sigma$ value will be calculated. This is the standard deviation of the noise required for this query. To answer query $Q_{m}$ with a sensitivity of $\Delta Q_{m}$, a zero-mean Gaussian noise with standard deviation $\sigma = \sqrt{2ln\frac{1.25}{\delta_{m}}} \times \frac{\Delta Q_{m}}{\epsilon_{m}}$ is added to the true query result. Since Gaussian noise can be calculated from the standard deviation, the proposed system stores the standard deviation and uses it as a basis of comparison for reusing noise. The system retrieves all previous transactions from the blockchain and perform comparison. It first checks the blockchain for any existing record with the same query type and standard deviation used. If an existing record is found in the blockchain, the algorithm will return the same result as the output of differential privacy. If no existing record is found, it will compare the standard deviation of a new query with previous queries made. The algorithm reuses Gaussian noise by injecting noise to the previous noise to generate a new noise that fulfils the privacy requirements. As such, full reuse will be possible if the new standard deviation is larger than the minimum standard deviation of all previous queries from the same type. The algorithm will then calculate the new noise that needs to be added to the previous results and derive the new result. If the new standard deviation is smaller than the minimum standard deviation previously generated, it will not be able to fully reuse any previous noise. The algorithm can only reuse a fraction of a previous noise and computes additional noise to be added. The query will then be forwarded to the server to add the computed noise to the partially reused noisy response.

\subsection{Output display}

\begin{figure}[h!]
\centering
\frame{\includegraphics[scale=1,width=\linewidth]{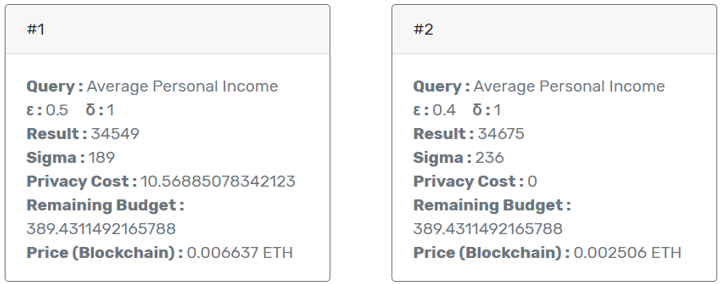}}
\caption{Displaying of output with $\epsilon$ privacy cost. }
\captionsetup{justification=centering}
\label{fig:outputdisplay}
\end{figure}

This demo displays the output of any query performed in a card format that is shown in Fig. \ref{fig:outputdisplay}. These cards will pop up once the output is available. The header of the card contains a query ID that is generated for the query submitted. In the body, it contains the query type that was requested, the noisy response generated by the application, standard deviation ($\sigma$) calculated, blockchain price, privacy cost, and the remaining privacy budget. The privacy cost displayed is $\epsilon$ privacy cost.

\section{Implementation}
\begin{figure}[h!]
\centering
\frame{\includegraphics[scale=1]{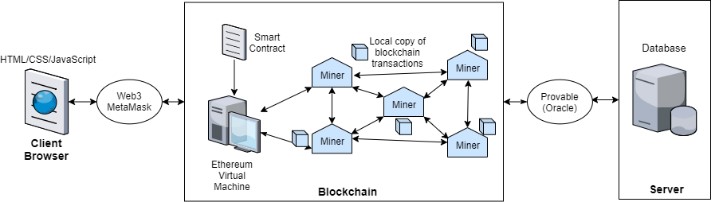}}
\caption{Overall software architecture.}
\captionsetup{justification=centering}
\label{fig:architecture}
\end{figure}

The proposed system is developed by making use of Bootstrap, Ethereum, MetaMask, Web3.js, Truffle Suite, Provable, MongoDB and Heroku. Fig. \ref{fig:architecture} shows the architectural diagram of the proposed system.

The DApp consists of the frontend client browser and the backend that runs on a decentralised platform, Ethereum and MongoDB database at the hosted server. Truffle Suite and Provable is used for the development of the smart contract. For the frontend client browser, index.html and app.css files define the webpage displays for the user. It also makes use of Bootstrap for responsive interactions with the user. The frontend also contains a app.js file that maps items from index.html, interacts with Web3.js, processes calculations and parses it to the blockchain.

\section{Conclusions}
In this paper, a simple DApp demo is developed to illustrate the use of blockchain to reuse previously generated noisy responses from differential privacy algorithm. In the future, this demo can be improved with the addition of graphical elements to better show the effects of the system such as the increase in number of queries that user can submit without exceeding the privacy budget. The demo can also be further improved by analyzing and quantifying the extent of privacy preservation with the reuse of noise.

\Urlmuskip=0mu plus 1mu\relax
\bibliographystyle{ACM-Reference-Format}
\bibliography{references}

\end{document}